# 1. Title page

**Influence of force-length relationship and task-specific constraints on finger force-generating capacities**

## Running head title

Force-length relationship and grip strength

## Authors

GOISLARD DE MONSABERT, Benjamin[1]; CAUMES, Mathieu [1]; BERTON, Eric[1]; VIGOUROUX, Laurent[1].

## Affiliations

[1]Aix-Marseille University, CNRS, ISM, Marseille, France

## Correspondence

Benjamin GOISLARD DE MONSABERT

Email : benjamin.goislard-de-monsabert@univ-amu.fr

Address: UMR 7287 CNRS & Aix-Marseille Université / Faculty of Sport Science, CP 910 / 163, av. de Luminy / F-13288 Marseille cedex 09 (FRANCE)

Phone: +33 (0)4 13 34 59 21

Fax: +33 (0)4 91 17 22 52

## 2. Abstract and key terms

Grip strength loss in extended and flexed wrist postures has been explained by reduced force-generating capacities of extrinsic finger flexor resulting from non-optimal length, owing to the force-length relationship. Recent works suggested that other muscles, especially wrist extensors, participate in this grip strength loss. The objective of this study was to clarify the role of the force-length relationship in finger force production. 18 participants performed maximal isometric finger force production during pinch grip (Pinch) and four-finger pressing (Press) tasks in four different wrist postures (extended, flexed, neutral, spontaneous). The maximum finger force (MFF), finger and wrist joint angles as well as activation of four muscles were determined using dynamometry, motion capture and electromyography. The force and length of the four muscles were estimated from joint angles and muscle activation using a musculoskeletal model. MFF decreased for flexed wrist during Pinch but remained stable across wrist postures during Press. The results suggested that the loss of pinch grip force in deviated wrist posture is partially related to force-length relationship of finger extensors. In opposition, maximal finger force during Press was not affected by the modulation of muscle capacities and was rather associated to finger interdependence resulting from mechanical or neural factors.

## 3. Introduction

Human hand is essential to our daily life activities thanks to a rich anatomy that allows adaptation to a wide variety of situations and objects. Many specificities enable this adaptation, such as the thumb opposition or the abundance of joint degrees of freedom (DoF). Among these specificities, the wrist has a crucial role for grasping tasks. This joint allows orienting and positioning the hand in numerous configurations thanks to its wide range of motion. Furthermore, the wrist is biomechanically coupled to the finger actions since the hand musculature includes pluri-articular extrinsic muscles, such as *flexor digitorum superficialis* (FDS), originating in the forearm and inserting on the phalanxes, thus crossing multiple joints including the wrist. Consequently, the action of finger extrinsic flexors to grasp an object inherently induces mechanical actions at the wrist, thus placing the wrist mechanical equilibrium as a necessary condition of successful grip.

The hand/wrist chain mechanical equilibration relies on two biomechanical phenomena. The first one concerns the high co-contraction of finger and wrist extensors during prehensile tasks. When grasping an object, the extrinsic flexors generate the finger force to seize the object thus inducing a flexion moment at the wrist, that must be statically equilibrated to ensure a stable wrist position. Because the object is at mechanical equilibrium, this flexor moment is entirely compensated by wrist and finger extensors[22] with force levels sometimes equivalent to those of flexors[10]. For non-prehensile tasks, the wrist flexion moment induced by extrinsic flexors is nearly balanced by the reaction force of the surface, hence reducing extensor implication[22]. The second phenomenon of the hand/wrist biomechanics is the grip strength loss in deviated wrist postures. The maximum grip force indeed varies with wrist position for both power[19] and pinch grip tasks[5,11,14] following a bell-shaped curve with a maximum for neutral or slightly extended wrist and a decrease of up to 30% in flexed and extended position[5,11,14,19]. A proposed hypothesis was that grip force losses at non-optimal wrist positions were due to a decrease in the force-generating capacities of the extrinsic finger flexors, main agonists of the task, owing to the force-length relationship. This relationship describes the relationship between the maximum force a muscle can produce and its length[20]. The force-generating capacity of a muscle is maximal for an optimal length and decreases when the muscle shortens or lengthens from that optimal point, thus resembling the evolution of maximum grip force against wrist posture. Because of this similarity, it was hypothesised that a modification of wrist position might result in a change of finger flexor muscle length, inducing a loss of muscle force capacity and ultimately a decrease in grip force capacity. Despite being generally admitted, this hypothesis has rarely been investigated with quantified data at the muscle level. Furthermore, considering the high co-contraction levels during prehensile tasks, it is reasonable to assume that finger and wrist extensors could play a role in grip strength variations.

The lack of proof of regarding the role of force-length relationships in grip force loss in deviated wrist positions is due to a lack of *in vivo* data of finger muscle mechanics, since most studies focused on the larger muscles, from the upper [16] and lower[2,27] limbs. Thanks to recent studies on hand muscle force-length relationships[8,12], new data were provided and showed that during a power grip task the capacities of finger flexors remained close to optimal despite changes of wrist postures[3], while the wrist extensor capacities decreased importantly for flexed and extended wrist. Those results suggest that wrist extensors could thus be seen as the muscle group inducing a large part of grip strength variations, contrary to previous hypothesis attributing it to the finger flexors. This hypothesis was further supported by the fact that the optimal wrist posture maximising grip force corresponded to the most optimal length configuration for all muscles. Nevertheless, such hypothesis was only observed for a power grip task and thus need to be explored in other hand force production configuration such as pinch grip and non-prehensile tasks. Pinch grip is indeed also exposed to grip force loss with deviated wrist positions[5,11,14] and is among the most common used grip[25]. The influence of wrist posture on finger forces during non-prehensile tasks, such as four-finger pressing, is less studied whereas those tasks offer an interesting paradigm, with lesser co-contraction of wrist extensors[22] and specific wrist equilibrium constraints due to finger force sharing[17]. Studying those other hand force production tasks appears thus relevant to verify the conclusions established with power grip and provide new insight into factors affecting grip strength.

The present study will explore the role of the force-length relationship in maximal finger force production and muscle coordination in two tasks with different biomechanical wrist constraints: a thumb-index pulp pinch grip task (Pinch) and a four-finger pressing (Press) task. A setup was designed to measure finger force, hand and wrist kinematics and muscle activity for different wrist postures. A previously developed musculoskeletal model[3] was used to explore the muscular coordination in relation with finger forces. This model estimates the force and length of four muscles representative of the main muscle groups for grasping (FDS; *flexor carpi radialis*, FCR; *extensor carpi radialis*, ECR; *extensor digitorum communis*, EDC) using *in vivo* electromyography (EMG) and kinematics measurements as input. Three hypotheses are formulated based on the hand musculoskeletal biomechanical functioning and on our previous works. Because prehensile task involves higher co-contraction of extensors[22] and thus more muscles, the first hypothesis is that maximal finger force will vary more with wrist posture during Pinch compared to Press. Since flexor muscle capacities are less impacted by wrist posture[3,8], our second hypothesis was that finger and wrist extensor capacities are importantly correlated with the grip modulation. Finally, as observed in our previous study on power grip[3], the third hypothesis was that the length configuration of all muscles must be considered to explain the wrist posture that maximises finger force.

## 4. Materials & Methods

### Participants

Eighteen volunteers (9 men & 9 women) that did not suffer from any musculoskeletal disorders within the last six months participated in the study. Before starting the experiment, participants signed an informed consent. The protocol was approved by a national ethics committee (CERSTAPS).

Anthropometric measurements were taken for each participant and included both hand length and reference muscle-tendon unit (MTU) lengths ($L_r^{mtu}$) (Table 1) of the 4 muscles considered in this study: ECR, EDC, FCR, and FDS.

### Protocol

Participants had to press maximally with their fingers on a surface rigidly attached to a force sensor by using either a pinch grip (Pinch) or a four-finger pressing (Press) technique (Figure 1). During Pinch, participants exerted force using the index and thumb fingertip pulp. During Press, participants exerted force on a surface with the pulp of the 4 long fingers. Participants waited for a verbal cue from the experimenter, then reached the maximal force as fast as possible and maintained it for 5 seconds. Participants were standing with the shoulder at about 30°of flexion and adduction, the elbow at 45°of flexion and the index finger at 30°for both metacarpophalangeal (MCP) and proximal (PIPi) and distal (DIPi) interphalangeal and joints. Four wrist postures were tested. Three postures were imposed: a neutral (WrN) position with the wrist at 0° of flexion/extension and a flexed (WrF) and an extended (WrE) posture, at 20° away from the maximum angle reachable by the participant. For the fourth posture, called spontaneous (WrS), the participant freely chose the wrist angle to perform the task. Participants performed in total 16 isometric finger force production tasks (2 tasks x 4 wrist postures x 2 trials).

*Insert Figure 1 around here*

Before beginning the Pinch and Press task, the participant performed seven maximum voluntary contraction (MVC) to normalize EMG levels. During all the MVC tasks, the participant was sitting, holding a handle (3.5cm in diameter) and respected a shoulder and elbow posture as described above and a neutral wrist posture. Four tasks consisted in exerting wrist moments in flexion, extension, ulnar and radial deviation. One task consisted in extending finger. The last task consisted in a maximal power grip (finger flexion).

Trials were blocked-randomized such that participants always began with the MVC trials and performed finger force trials in second, but trials were randomized within each session. Each trial was separated by a two-minute rest to avoid muscle fatigue.

### Data Acquisition

A 6-axis force and torque sensor (Nano25, ATI, Apex, NC, 2000Hz) was used to measure the force applied by the fingers. The sensor width was 5.5 cm, corresponding the average size maximising pinch grip strength[6]. It was mounted on a height-adjustable pole to adapt to participants' morphology.

A motion capture system (Qualysis, Göteborg, Sweden; 100 Hz) composed of six cameras was used to track 13 markers with an 8-mm radius placed on anatomical landmarks of the hand and wrist. The marker placement followed those of the previous study[3] and are detailed in Electronic Supplementary Materials (ESM1).

A wireless EMG system (Trigno, Delsys, Natick, MA, 2000 Hz) was used to record the activities of 10 muscles including the four muscles of the musculoskeletal model (ECR, EDC, FCR, FDS). Two ulnar deviators (*extensor* and *flexor carpi ulnaris*), two elbow (*biceps brachii*, *triceps*) and two shoulder (*anterior* and *posterior deltoïd*) muscles were also monitored. Before placing electrodes, the skin was shaved, sanded and rinsed with an alcoholic solution. The electrode placement for FDS, EDC, FCR and ECR followed the previous study[3] whereas other recommendations[28] were used for other muscles. Functional contractions, targeting each muscle, were used to minimise crosstalk[9].

The EMG and force sensor signals were synchronously recorded with the kinematics data via an analogue card interfaced with the Qualysis system.

### Data processing

The six signals from the force sensor were low-pass filtered (Butterworth, 10Hz, order 2, zero-phase) and converted to force components via a calibration matrix. The finger force was computed as the norm of the three force components. The maximum finger force (MFF) corresponded to the mean of the finger force on a 500-ms window centred on the force peak. For each task, the normalized MFF (nMFF) was calculated by dividing the MFF determined in the current trial by the maximal MFF value among all trials of the current task.

The marker coordinates were low-pass filtered (Butterworth, 10Hz, order 2, zero-phase) and averaged on the same 500-ms window as the force to determine joint angles ($\theta^j$). The wrist and MCPi joints were described by two DoF in flexion/extension and radial/ulnar deviation and the DIPi and PIPi joints by one DoF in flexion/extension. For each joint, a distal and a proximal segment coordinate system were calculated from marker positions (see Electronic

Supplementary Material ESM1) and joint angles were extracted from the relative orientation matrix using a Z-Y-X (flexion/abduction/pronation) sequence of Cardan angles[10]. Flexion, and radial deviation angle were considered positive.

EMG signals were bandpass filtered (Butterworth, 10-400Hz, order 2, zero-phase). Then, the root mean square (RMS) was calculated for each signal on the same 500-ms window as the force. The muscle activation ($a^m$) of each muscle was calculated by normalizing the RMS value in a trial by the maximal RMS value observed among all trials, both MVC and finger force exertion, for that muscle.

The same musculoskeletal model as the study focusing on power grip[3] was used to estimate the force ($F^m$) and length ($L^m$) of FDS, EDC, FCR and ECR from measured anthropometrics ($L_r^{mtu}$, $L_{hand}$), joint angles ($\theta^j$) and muscle activation ($a^m$). This model is mainly based on experimentally-derived Force-Length-Activation relationship obtained from previous studies[8,12] and is presented in detail in Electronic Supplementary Materials (ESM2). Those relationships consider the activation-dependency of the muscle contraction, especially the shift in optimal length at low activation. $F^m$ and $L^m$ were normalized by maximal isometric force $F^{max}$ and optimal muscle length $L_{opt}$ at maximal activation ($a^m$=1), respectively.

As in our previous study[3], a criterion ($C_{\Delta l}$) based on muscle length was computed :

$$C_{\Delta l} = \sum_{m=1}^{4} \frac{|L^m - L_{opt}|}{L_{opt}} \qquad \text{Equation 1}$$

This criterion consisted of the sum across all muscles of the normalised absolute difference between the current length ($L^m$) and the optimum length ($L_0$). A low criterion value indicate that all muscles are close to their optimum length and hence that the overall force-generating capacities are maximised.

## Data Analysis

To test our first hypothesis related to the influence wrist position and the type of task, two-way repeated ANOVA was conducted for each variable ($\theta^j$, MFF, $a^m$, $F^m$ and $L^m$) to evaluate the effects of *posture* (WrE, WrS, WrN, WrF) and *task* (Pinch, Press) with a significance level of *p*=0.05. If the effect was significant, differences between conditions were evaluated using multiple pairwise *t*-test comparisons with an adjusted significance level (*p_{adj}*) using Bonferroni correction. Our second hypothesis aiming to identify muscles explaining finger force variations against wrist posture was tested using multiple regression analyses comparing estimations of MFF from all possible muscle force combinations. All combinations across the four muscles were tested (4 single, 6 pairs, 4 trios, and all four). For each combination, a regression model was estimated, and the associated corrected Akaike Information Criterion (AICc)[1] was computed. The AICc score allows to characterize the plausibility of each combination to explain the MFF variations with a penalty term on the number of muscles. For clarity, the combinations were compared using the ΔAICc consisting in the difference between the AICc and the best AICc value. A muscle combination with a ΔAICc value less than 2 is considered substantial, between 2 and 4 is less plausible, and above 4 is implausible. In addition, the level of explained MFF variations of each muscle combinations was computed through the adjusted R-square ($R^2_{adj}$). All statistical analysis were made with R Statistical Software (v4.1.1; R Core Team 2021). To test our third hypothesis, the muscle length criterion value was computed for all trials of all participants and plotted against wrist flexion/extension angle to evaluate which posture maximised muscle force-generating capacities.

## 5. Results

### Measured joint Angles

All joint angles are presented in Figure 2. Wrist flexion/extension varied from -67.8±14.0 degrees for Press in WrE to 37.8±13.6 for Press in WrF. The two-way ANOVA showed a significant effect of *posture* ($F(3,51)=406.2$; $p=1.42\times10^{-35}$) and no significant effect of *task* ($F(1,17)=13.5$; $p=0.64$) and of *posture*×*task* interaction ($F(2.05,34.5)=2.1$; $p=0.14$). The wrist flexion increased progressively from extension to flexion with each posture being significantly different from the others ($p_{adj}<10^{-8}$).

Wrist radial/ulnar deviation varied from 2.3±9.2 degrees during Press in WrE to 13.5±8.2 during Pinch in WrN. The effect of *posture* ($F(3,51)=16.9$; $p=9.6\times10^{-8}$) was significant but the effect of *task* ($F(1,17)=1.6$; $p=0.19$) and *posture*×*task* interaction ($F(3,51)=1.0$; $p=0.40$) were not. The wrist radial deviation from WrE to WrN and ulnarly from WrN to WrF with each posture being significantly different from the others ($p_{adj}<10^{-8}$), except between WrS and WrF ($p_{adj}=1.0$).

MCPi flexion/extension varied from 7.8±18.0 degrees during Press in WrE to 29.1±13.9 during Press in WrF. The effect of *posture* ($F(3,51)=6.1$; $p=0.001$), *task* ($F(1,17)=7.5$; $p=0.014$) and *posture*×*task* interaction ($F(3,51)=12.3$; $p=3.5\times10^{-6}$) were all significant. During Pinch, MCPi flexion remained was not different between postures ($p_{adj}>0.07$). During Press, MCPi flexion was also not different between postures ($p_{adj}>0.076$), except in WrF where flexion was higher than in WrN and WrE ($p_{adj}<0.010$). MCPi flexion was slightly higher during Pinch than during Press ($p_{adj}<0.013$), except in Flexion ($p_{adj}<0.013$).

MCPi radial/ulnar deviation varied from −3.1±8.3 degrees during Pinch in WrE to 10.9±7.7 during Pinch in Flexion. The effect of *posture* ($F(3,51)=21.5$; $p=3.9\times10^{-9}$) and *task* ($F(1,17)=50$; $p=1.9\times10^{-6}$) were significant but the effect *posture*×*task* interaction was not ($F(3,51)=1.9$; $p=0.15$). MCPi was more ulnarly deviated during Press ($p_{adj}=3.2\times10^{-16}$). Its value increased

*Insert Figure 2 around here*

progressively from WrE to WrF with all pairs of postures being different ($p_{adj}$ <0.034), except between WrS and WrN ($p_{adj}$ =1.0).

PIP flexion/extension varied from 20.7±11.8 degrees during Press in WrE to 38.4±8.9 during Pinch in WrE. The effect of *task* (F(1,17)=25.8; $p$=9.3×10$^{-5}$) and of *posture*×*task* interaction (F(3,51)=5.3; $p$ =0.003) were significant but the effect of *posture* (F(3,51)=0.2; $p$=0.92) was not. PIP flexion was higher during Pinch at each posture ($p_{adj}$<0.047), expect in WrF ($p_{adj}$=0.457).

DIP flexion/extension varied from 25.8±20.9 degrees during Press in WrF to 41.6±9.8 in Pinch in WrS. The effect of *task* (F(1,17)=8.7; p=0.009) was significant but the effect of *posture* (F(1.67,28.42)=1.4; $p$=0.26) and *posture*×*task* was not (F(2.10,35.66)=2.6; $p$=0.087). DIP angle was higher in Pinch ($p_{adj}$ =0.001)

## Measured maximum finger force

The nMFF varied from 0.98±0.03 during Pinch in WrS to 0.71±0.22 during Press in WrF (Figure 3). The two-way ANOVA showed a significant effect of both *posture* (F(3,45)=8.0; $p$=2.3×10$^{-4}$), *task* F(1,15)=6.5; $p$=0.0022) and *posture*×*task* interaction (F(3,51)=3.8; $p$=0.017). During Pinch, nMFF was lower in WrF compared to the other postures ($p_{adj}$ < 0.011). During Press, nMFF was only lower in WrE compared to WrF ($p_{adj}$=0.017). The MFF was higher during Pinch for both WrE and WrS postures ($p_{adj}$=0.018).

## Measured muscle activation

Muscle activations are presented on Figure 4. For FDS, the two-way ANOVA showed a significant effect of *posture* (F(3,45)=4.4; $p$=0.29), but not of the *task* (F(1,15)=0.67; $p$=9.7×10$^{-7}$) nor the *posture*×*task* interaction (F(3,45)=0.55 ; $p$=0.02). FDS activation was the same across all postures ($p_{adj}$>0.45), except in WrE with lower levels compared to all other postures ($p_{adj}$<0.004).

For FCR, the effect of *posture* (F(3,45)=6.0; $p$=0.001), *task* (F(1,15)=43.8; $p$=8.2×10$^{-6}$) and the *posture*×*task* interaction (F(3,45)=3.7; $p$=0.018) were all significant. FCR activation was higher during Press ($p_{adj}$=3.2×10$^{-13}$), remain the same across postures ($p_{adj}$>0.05) except in WrE showing a lower value than WrS ($p_{adj}$=0.08).

For EDC, the effect of *task* (F(1,15)=183; $p$=8.4×10$^{-10}$) and *posture* (F(3,45)=17.7; $p$=1×10$^{-7}$) were significant but the *posture*×*task* interaction was not (F(3,45)=4.4; $p$=0.061). EDC activation was higher during Pinch ($p_{adj}$=1.2×10$^{-29}$) and decreased progressively from WrE to WrF, with most postures being different from the others ($p_{adj}$<0.004), except WrN that was not different from WrS and WrF ($p_{adj}$>0.2).

For ECR, the *task* (F(1,15)=62.7; $p$=9.8×10$^{-7}$) and of the *posture×task* (F(3,45)=4.4; $p$=0.02) interaction were significant but the *posture* was not (F(3,45)=1.3; $p$=0.29). ECR activation was higher during Pinch ($p_{adj}$=9.4×10$^{-18}$).

## Estimated muscle force and length

The normalised muscle forces estimated using the musculoskeletal model are presented in Figure 4. For FDS, the two-way ANOVA showed a significant effect of *posture* (F(3,45)=8.9; $p$=9.3×10$^{-5}$), but not of the *task* (F(1,15)=0.67; $p$=0.73) nor the *posture×task* interaction (F(3,45)=0.55; $p$=0.26). FDS force was the same across all postures ($p_{adj}$>0.54), except in WrE with lower levels compared to all other postures ($p_{adj}$<0.002).

For FCR, the effect of *posture* (F(3,45)=29.1; $p$=1.3×10$^{-10}$), *task* (F(1,15)=36.5; $p$=2.3×10$^{-5}$) and the *posture×task* interaction (F(3,45) =8.7; $p$=0.0001) were all significant. FCR force was higher during Press ($p_{adj}$=1.2×10$^{-10}$) and showed lower levels in WrE and WrF ($p_{adj}$<0.0001) compared to WrN and WrS which were not different ($p_{adj}$=0.33).

For EDC, the effect of *task* (F(1,15)=175; $p$=1.1×10$^{-9}$) and *posture* (F(3,45)=15.2; $p$=5.9×10$^{-7}$) were significant but the *posture×task* interaction was not (F(3,45)=2.6; $p$=0.06). EDC activation was higher during Pinch ($p_{adj}$=2.8×10$^{-29}$) and showed lower levels in WrE and WrF ($p_{adj}$<0.0005) compared to WrN and WrS which were not different ($p_{adj}$=0.98).

For ECR, the *task* (F(1,15)=58.4; $p$=1.5×10$^{-6}$), *posture* (F(3,45)=11.4; $p$=1.1×10$^{-5}$) and the *posture×task* (F(3,45)=3.9; $p$=0.015) interaction were all significant. ECR activation was higher during Pinch ($p_{adj}$=6.5×10$^{-16}$) and showed lower levels in WrE and WrF ($p_{adj}$<0.001) compared to WrN and WrS which were not different ($p_{adj}$=0.77).

The normalised muscle length estimated using the musculoskeletal model are presented in Figure 4. For FDS, the two-way ANOVA showed a significant effect of *posture* (F(3,45)=154; $p$=5.5×10$^{-17}$), *task* (F(1,15)=25.6; $p$=0.0001) and *posture×task* interaction (F(3,45)=15.5; $p$=1.7×10$^{-5}$). FDS length decreased from WrE to WrF with all pairs of postures being different ($p_{adj}$<0.001) and was slightly higher during Press than Pinch, only in WrE ($p_{adj}$=2.1×10$^{-5}$).

For FCR, the effect of *posture* (F(3,45)=256; $p$=2.7×10$^{-28}$) was significant but the effect of *task* (F(1,15)=0.02; $p$=0.88) and *posture×task* interaction (F(3,45)=1.5; $p$=0.22) were not. FCR length decreased from WrE to WrF with all pairs of postures being different ($p_{adj}$<5×10$^{-7}$).

For EDC, the effect of *task* (F(1,15)=15; $p$=0.016) and *posture* (F(3,45)=226; $p$=6.79×10$^{-15}$) were significant but the *posture×task* interaction was not (F(3,45)=0.518; $p$=0.55). EDC length increased from WrE to WrF with all pairs of postures being different ($p_{adj}$<5×10$^{-5}$) and was slightly higher during Pinch ($p_{adj}$=0.038).

For ECR, the effect of *posture* (F(3,45)=242; *p*=8.8×10$^{-28}$) was significant but the effect of *task* (F(1,15)=0.159; *p*=0.69) and *posture*×*task* interaction (F(3,45)=187; *p*=0.18) were not. ECR length increased from WrE to WrF with all pairs of postures being different ($p_{adj}$<5×10$^{-6}$).

## Multiple regression analysis

The multiple regression analysis results between MFF and muscle forces are presented in Table 2. During Pinch, the most plausible combination of muscle forces explaining MFF variations was the one including those of both extensors, i.e., ECR and EDC. The six best combinations (ΔAIC<4) included ECR while the three worst included only flexors. During Press, the finger flexor force, i.e., FDS, alone was the most plausible among all tested combinations to explain MFF variations The seven best combinations (ΔAIC<4) included FDS while the three worst included only extensors.

*Insert Table 2 around here*

## Muscle length criterion

The muscle length criterion evolution is presented on Figure 5 and was minimal during WrN (0.09±0.03 during Pinch and 0.06±0.03 during Press) and maximal for WrE (0.24±0.03 during Pinch and 0.30±0.06 during Press). Intermediate values were observed during WrS (0.12±0.05 during Pinch and 0.13±0.07 during Press) and WrF (0.14±0.03 during Pinch and 0.14±0.05 during Press)

*Insert Figure 5 around here*

# 6. Discussion

The objective of this study was to explore the influence of the muscle force-length mechanics, modulated with different wrist postures, on finger force during two force production tasks. An experimental protocol was developed to synchronously measure finger force, joint angles, and muscle activation during a prehensile (Pinch) and non-prehensile (Press) task performed in four wrist postures. Conjointly, a previously developed musculoskeletal model[3] provided an estimation of the force and length of four muscles representative of four of the main hand muscle groups based on measured joint angles and muscle activations.

The wrist postures were different from each other and remained the same across tasks, despite some variations in finger posture (Figure 2). Wrist deviation remained stable with only slight differences (<10° variations) when the wrist moved in flexion. The finger joint angles remained also stable against the different wrist postures but varied significantly between tasks, especially for MCPi that was more flexed and ulnarly deviated during pinch grip. These differences were due to joint angle adjustments to reach a maximal performance during force exertions. Thanks to the instructions and visual control before force exertion, those natural posture variations remained low (below 15°) compared to those of imposed wrist postures (above 100°). It can be thus considered that variations of finger force and muscle mechanical variables described below were thus mainly influenced by the modulation of imposed wrist postures.

The influence of wrist posture on finger force production differed between the two tasks with larger variations during Pinch than Press, confirming our first hypothesis (Figure 3). During Pinch, a flexed wrist induced about 25% loss of strength compared to other postures while extension did not significantly modify the force capacity. This dissymmetry was also observed in the literature, with higher loss in flexed than extended wrist [5,11,14]. Authors have mainly hypothesised that the decreased pinch grip strength in wrist flexion was induced by the force-length relationships of finger extrinsic flexor, main agonists of the gripping actions. A flexed wrist indeed results in shorter fibre lengths for finger extrinsic flexors and thus reduces their capacities [11,14] but this was never confirmed with quantified data. Furthermore, as suggested by our previous study[3], the sole study of finger flexor capacity might not be sufficient to explain grip force variations. During Press, the total finger force exerted by the four fingers did not vary across wrist postures. This resultant force stability was also observed in a previous study[13] and might be the results of motor control constraints, i.e., the finger force deficit[17]. During multi-finger maximal force exertion, each finger indeed produces less force than when maximally exerting force alone, possibly because of finger interdependence phenomenon related to mechanical and nervous factors[21]. The mechanical constraints are related to anatomical structures, e.g., extrinsic muscles spread in different compartments to control each finger, or to task demands, e.g., the finger force sharing requires balancing a secondary moment induced by fingertip force, in prono-supination here. Nervous constraints have been identified at the peripheral level, e.g. synchronous firing of motor units from different extrinsic muscle compartments, or at the central level, overlap of the cortical territories associated with adjacent digits. Our results thus suggests that finger force-generating capacity during multi-finger force pressing tasks is not limited by force-length constraints but rather by finger interdependence arising from anatomical interconnections, task demands or neural restrictions. Consequently, the modulation of finger extrinsic muscle lengths caused by the different wrist postures does not have the same effect depending on the task.

The muscle forces estimated by the musculoskeletal model (Figure 4) were lower in flexed and extended wrist postures compared to neutral and spontaneous ones, except for the extrinsic finger flexors (FDS) which remained stable. This result was mainly explained by the fact that the muscle length in flexion and extension were the furthest from the optimal length (dashed line on Figure 4). Because of these non-optimal configurations, the muscle force capacities for both extensors (ECR and EDC) and for the wrist flexor (FCR) were lower in most deviated postures, i.e., WrF and WrE. The stability of FDS muscle force against wrist posture variations is in agreement with results of previous studies[8,12], that showed this muscle remained on the optimal region, i.e., plateau, of its force-length relationship despite different wrist position. This results nevertheless seems in contradiction with the common hypothesis that the grip force loss in extreme wrist position is explained by reduced extrinsic finger flexor capacity. In our study, a decrease in maximum pinch grip force was observed for a flexed wrist whereas FDS muscle force was the same as in neutral and spontaneous postures. On the contrary, the major FDS force decrease was observed in wrist extension whereas the pinch force was not different than in neutral or spontaneous postures. Those observations tend to corroborate the results of our previous study on power grip[8] that the decrease in grip force with extreme postures cannot be solely explained by variations of FDS force-generating capacity and that other muscles participate in the phenomenon.

The estimated muscle forces varied importantly between the two tasks (Figure 4) with the higher levels of extensors (EDC and ECR) during Pinch and of the wrist flexor (FCR) during Press, except for FDS that participated equally in both tasks. Those differences are mainly due to task-specific muscle coordination observable at the EMG muscle activation levels (Figure 4). The estimated muscle length is indeed nearly identical between the two tasks (Figure 4), which was expected since wrist postures were equivalent. On the contrary, the differences in muscle activations between the tasks demonstrate the same trend as muscle forces. Those task-specific muscle coordination are consistent with the wrist mechanical equilibrium constraints[4,22]. During Pinch, because the forces applied by the fingers are balanced, high co-contraction levels of both finger and wrist extensors are required to balance the wrist flexion moment induced by the extrinsic finger flexors, main agonists of grip force exertion. Compared to the study on power grip[3], the implication of extrinsic finger extensor (EDC) is higher than for wrist extensors (ECR), probably traducing the need to stabilize finger joint which are not in contact with the object. During Press, the wrist flexion moment generated by finger flexors is balanced by the reaction force of the surface thus reducing the implication of extensors. Nevertheless, a high implication of wrist flexors (FCR) becomes necessary to stabilize the wrist. Those results showed that task-specific constraints influence the muscle coordination which in turns might modulate the role of the force-length relationship in hand force production.

The multiple regression analysis identified that muscle groups best explaining finger force variations against wrist postures varied between tasks. During Pinch, the three combinations of muscle forces best explaining finger force variations included both extensors, i.e., EDC and ECR (Table 2). This result is in line with the idea that antagonist muscles could potentially "drive" force loss against wrist posture during prehensile task, previously suggested during the power grip study[3]. As extensors are highly involved in grip tasks, their force-generating capacities seem to limit the global co-contraction levels reachable by the hand-wrist musculoskeletal system and ultimately reduce the ability to produce a finger force. This also could partially explain the fact that their tendons are frequently affected by lateral epicondylitis, also called tennis elbow[23]. If extensors are functioning close to their maximal capacities such that they limit the grip force, they could indeed be more at risk of overuse injuries, especially in forceful contexts with inadequate posture[23]. During Press, the relation between wrist posture and hand force production is unclear, mainly because the resultant finger force did not vary. Despite the results on the population did not show an effect of wrist postures on maximal finger force, the multiple regression analyses can inform us on the combination of muscle forces that can explain the variations within the population. From this point of view, the combination including both flexor muscle forces (FDS & FCR) best explained the finger force variations, confirming the performance in this task do not seem to be driven by the capacities of extensors. It should however be noted that the correlation coefficient showed relatively low values for both tasks ($R^2_{adj}<0.4$; Table 2). This is probably explained by the fact that only four muscles (FCR, FDS, EDC, ECR) were considered whereas over 30 muscles are actuating the five fingers and the wrist. The complete analysis of EMG activations (Electronic Supplementary Material ESM3) indeed showed that other wrist ulnar deviators were involved in both tasks. Nevertheless, to our knowledge, only the force-length relationships of the four considered muscles are available and assessing them required combining dynamometry, electromyography, motion capture and ultrasonography[8], resulting in lengthy protocols, i.e., 2h per participant for each muscle. Although the results above should be considered with caution, they tend to confirm our second hypothesis that the greater finger force loss for flexed wrist during Pinch is explained by loss finger extensor capacities.

Contrary to the results of our power grip study[3], the muscle-length criterion ($C_{\Delta l}$; Equation 1) is less related to maximum finger force variations (Figure 5). The criterion was minimal, suggesting high muscle force-generating capacities, for neutral wrist whereas performances were equivalent between neutral, extended, and spontaneous postures (Figure 2). This non-agreement between the muscle capacities and force performance was expected for Press where the MFF remained stable across all postures and confirms the limited effect of force-length relationship constraints in this task. As explained above, the reduction of muscle capacities with deviated wrist postures seems overridden by other task-specific constraints, such as finger force deficit[17]. During Pinch, the criterion suggested extended posture to be non-optimal in terms of force-generating capacities, but MFF was only reduced for a flexed wrist. This tends to confirm that the loss of Pinch strength in deviated wrist postures is partially related to force-length relationships, but that this link is not fully explained by our criterion. As already explained, only four muscles are considered such that the criterion only offers a partial overview of muscle capacities. Future studies should thus investigate the influence of other muscles such as finger intrinsic which play an important role in finger force production and joint stabilization during pinch grip[26].

Beyond those already mentioned, the results of this study present some limitations. First, the flexor digitorum profundus (FDP) was not considered in this study whereas it is the main agonist of fingertip force production since it flexes the most distal (DIP) joint whereas FDS is not. FDP activation however requires intra-muscular EMG and invasive electrodes making it rather difficult to use on both ethical and technical points of view. Nevertheless, studies have showed that FDS is equally implicated than FDP in pinch grip and fingertip force production[18,24,26] such that it provides a representative muscle of finger flexors during those tasks. Additionally, other factors could influence finger strength against wrist deviated postures other than active force-generating capacities. Deviated wrist postures also induce a contact between the extrinsic finger tendons and carpal bones or the retinaculum ligaments, resulting in tendon friction, altered force transmission or discomfort which were not studied here[11,14]. Such phenomenon was minimized by controlling that participant used wrist positions below their maximally reachable posture. Another limitation was that we did not measure the individual contribution of each finger in the Press task which. Although only scarce data is available for the effect of wrist posture during such task, a previous study showed that the finger force sharing was not modified by wrist flexion/extension[13]. Finally, the musculoskeletal model estimate muscle forces mostly from electromyography, i.e., forward approach, whereas most hand models use an inverse-dynamics approach relying external forces to determine individual muscle contribution. Considering the complexity of hand muscle coordination and the difficulties of investigating its small musculature, future studies could consider developing hybrid musculoskeletal models, already used at the lower limb[7].

This study provided new insights in the links between force production, posture, and muscle mechanics by combining a motion capture protocol with a musculoskeletal model. Although further studies are required, the results suggest that force-length relationship constraints influence finger force production in prehensile tasks but not in non-prehensile context. Other constraints, possibly at the nervous command level, could override the force-generating limitations. The data obtained is relevant for ergonomics of hand-object interaction to understand factors limiting strength and further improve hand tool design and musculoskeletal disorder prevention[15].

## 7. Funding and Conflicts of interests.

The authors certify that they have no Conflict of Interest in the subject matter or materials discussed in this manuscript and received no external funding for the work presented in this article.

## 9. Figures

Figure 1 – Illustrations of the experimental setup showing a participant equipped with reflective markers and electromyography electrodes during the Pinch task (a, b) and during the Press task (c, d). The lower panel picture (e) illustrates the force sensor and its support.

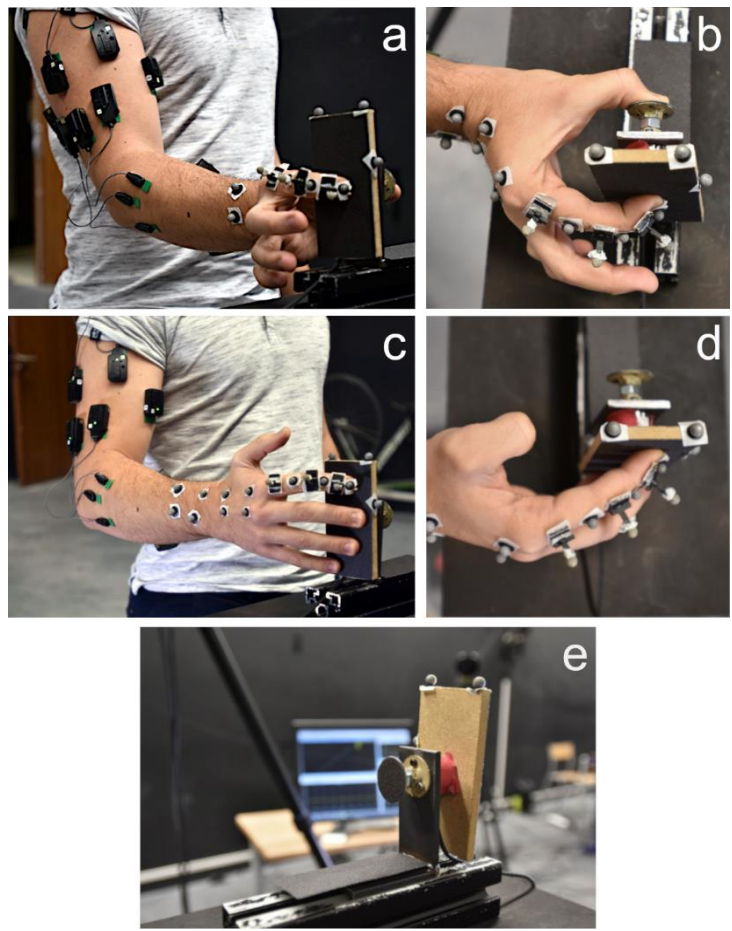

Figure 2 – Mean ± one standard deviation values of the measured joints angle ($\theta^j$) taken by the participants during the pinch grip (Pinch) and the four-finger pressing task (Press) in the different wrist postures, i.e., extension (WrE), spontaneous (WrS), neutral (WrN) and flexion (WrF). MCPi: index metacarpophalangeal joint; PIPi: proximal interphalangeal joint; DIPi distal interphalangeal joint. Solid grey lines represent values for the Pinch task and dashed black lines represent values for the Press task.

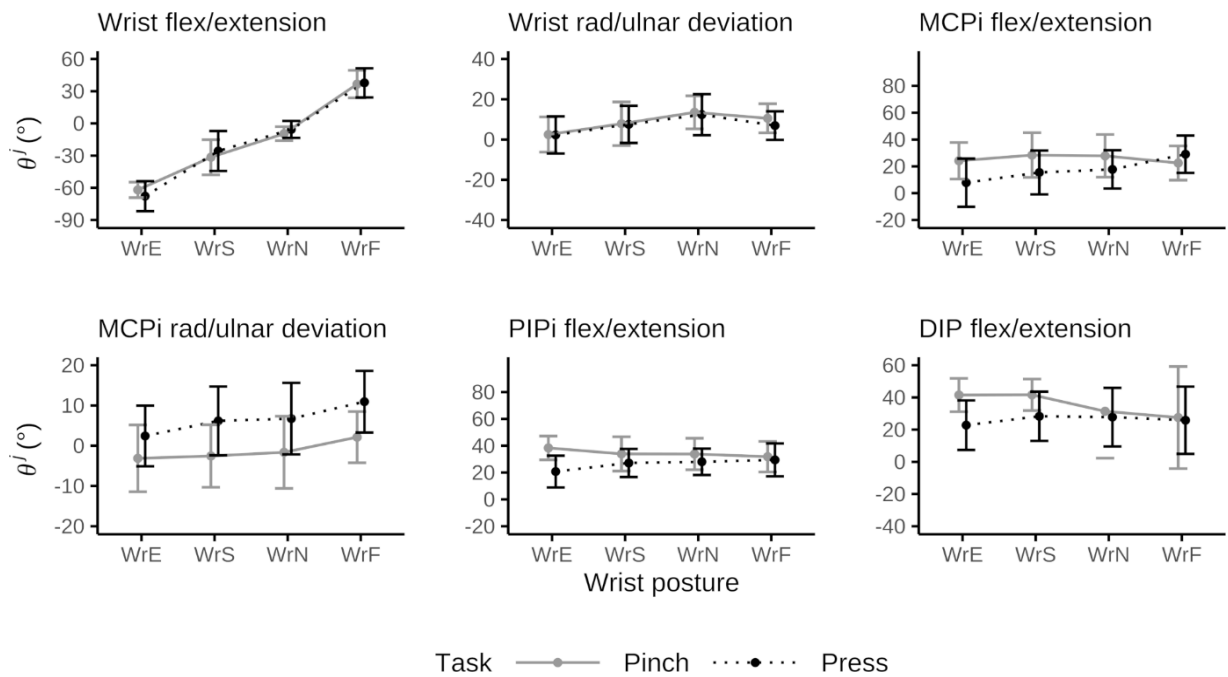

Figure 3 – Mean ± one standard deviation values of the normalized maximum finger force (nMFF) applied by the participants during the pinch grip (Pinch) and the four-finger pressing task (Press) in the different wrist postures, i.e., extension (WrE), spontaneous (WrS), neutral (WrN) and flexion (WrF). Solid grey lines represent values for the Pinch task and dashed black lines represent values for the Press task.

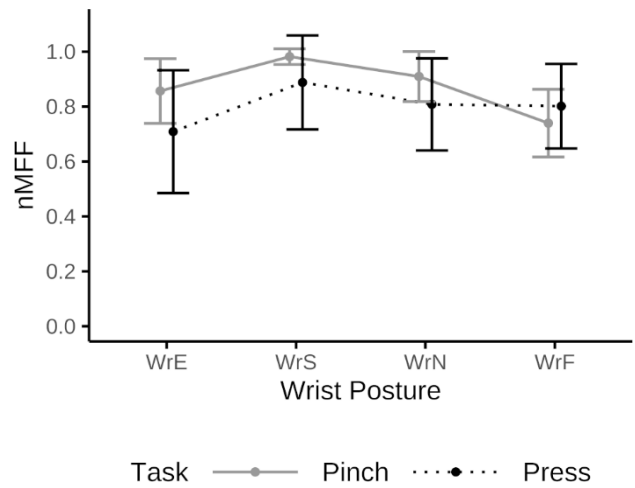

Figure 4 – Mean ± one standard deviation of measured muscle activation ($a^m$, upper panel) as well as normalized muscle forces ($F^m/F^{max}$; middle panel) and lengths ($L^m/L_0$; lower panel) estimated by the musculoskeletal model or the four muscles during the pinch grip (Pinch) and the four-finger pressing task (Press) in the different wrist postures, i.e., extension (WrE), spontaneous (WrS), neutral (WrN) and flexion (WrF). Solid grey lines represent values for the Pinch task and dashed black lines represent values for the Press task. The black dashed line on the lower panels indicates the value for which the muscle is at its optimal length. FCR: Flexor Carpi Radialis; ECR: Extensor Carpi Radialis; FDS: Flexor Digitorum Superficialis; EDC: Extensor Digitorum Communis.

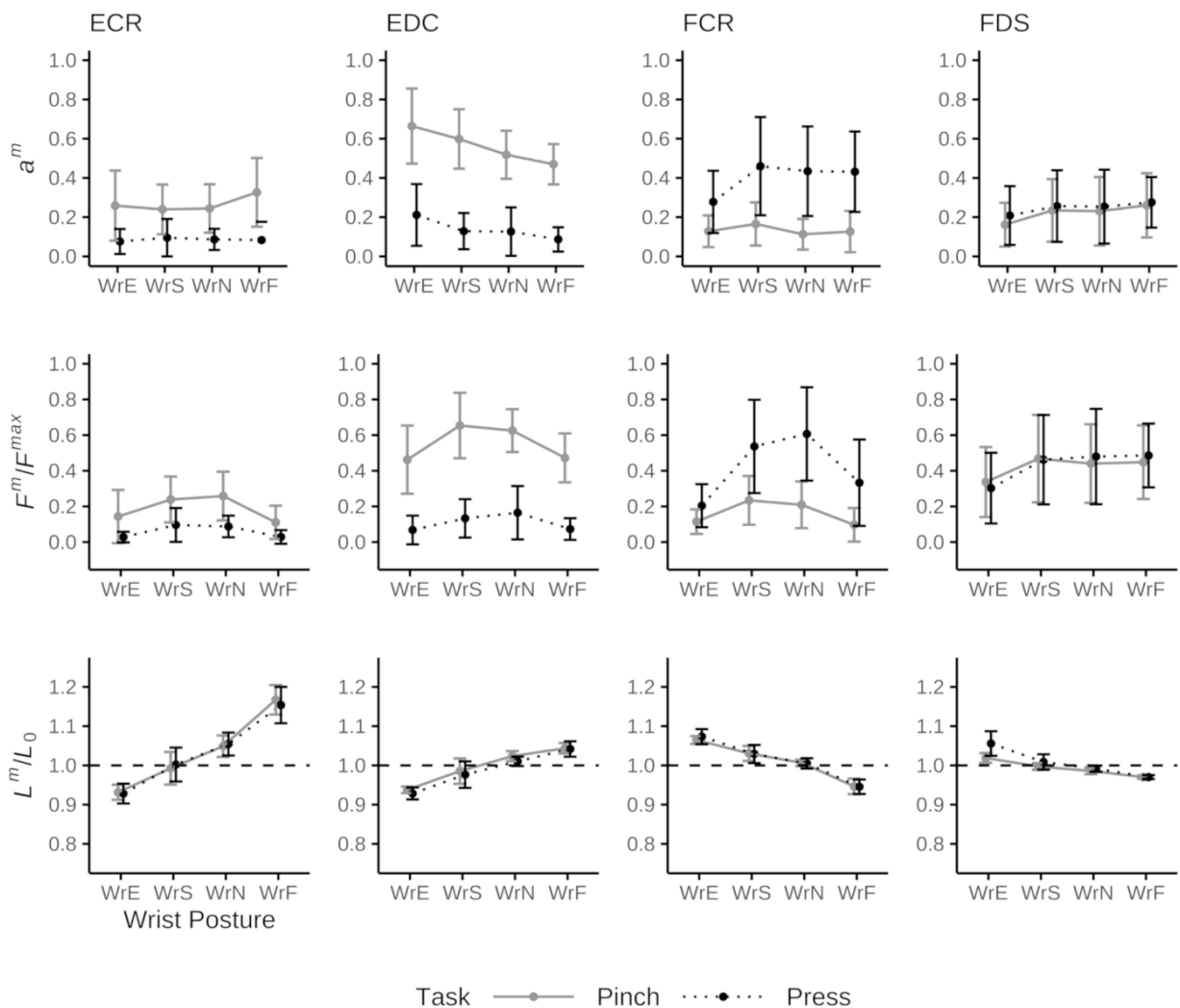

Figure 5 – Evolution of the muscle length criterion ($C_{\Delta L}$, Equation 1, upper panels) and the normalized maximum finger force (nMFF, lower panels) against wrist joint angle for the pinch grip (left panels) and the four-finger pressing task (right panels). Each point represents the value for a participant in one condition of wrist posture. A low $C_{\Delta L}$ value indicates that all muscles are close to their optimum length, hence that the overall force-generating capacities is maximal. Shaded areas and dashed lines represent the mean and standard deviation of the wrist joint angle in the spontaneous (WrS, dark) and neutral (WrN, light) postures. Positive value of the angle represents a flexion posture.

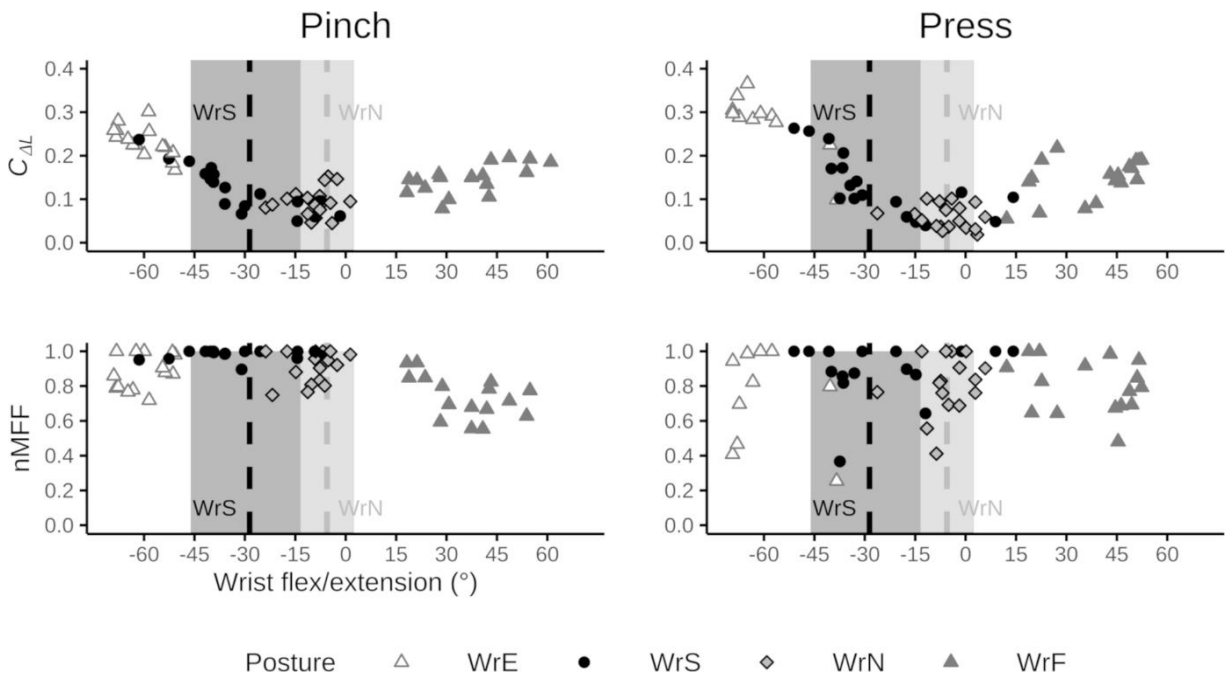

# 10. Tables

Table 1 – Mean anthropometric data measured for the population sample. $L_{hand}$ and $L_r^{mtu}$ corresponds to the hand length and reference muscle-tendon unit (MTU) length, respectively

|      | Height (cm) | $L_{hand}$ (cm) | $L_r^{mtu}$ (cm) | | | |
|------|-------------|-----------------|------|------|------|------|
|      |             |                 | EDC  | ECR  | FCR  | FDS  |
| Mean | 176.0       | 19.4            | 44.4 | 32.2 | 32.2 | 42.1 |
| *SD* | *9.1*       | *1.2*           | *3.1*| *2.6*| *2.3*| *3.2*|

Table 2 – Results of the multiple regression analysis aiming to identify muscle force combinations that best explain maximum finger force variations (MFF) using the corrected Akaike Information Criterion (ΔAICc). The combinations that are considered substantial (ΔAICc<2) are in bold font. Combinations are ranked according to the corrected Akaike Information Criterion (ΔAICc) and the combinations that are considered substantial (ΔAICc<2) are in bold font. $R^2_{adj}$ corresponds to the adjusted R-Squared that considers the number of variables included in the regression.

|  | Pinch | | | | Press | | |
|---|---|---|---|---|---|---|---|
| Muscle combination | Rank | ΔAICc | $R^2_{adj}$ | Muscle combination | Rank | ΔAICc | $R^2_{adj}$ |
| **ECR+EDC** | **1** | **0.00** | **0.26** | **FDS** | **1** | **0.0** | **0.36** |
| **ECR+EDC+FDS** | **2** | **0.47** | **0.10** | **FDS+FCR** | **2** | **1.2** | **0.36** |
| ECR+FDS | 3 | 2.10 | 0.17 | **FDS+ECR** | **3** | **1.5** | **0.36** |
| ECR+EDC+FCR | 4 | 2.11 | 0.08 | FDS+EDC | 4 | 2.2 | 0.35 |
| ECR+EDC FDS+FCR | 5 | 2.72 | 0.30 | FDS+FCR+ECR | 5 | 3.3 | 0.35 |
| ECR+FDS+FCR | 6 | 3.92 | 0.26 | FDS+FCR+ECR | 6 | 3.5 | 0.35 |
| ECR | 7 | 4.79 | 0.32 | FDS+ECR+EDC | 7 | 3.8 | 0.35 |
| ECR+FCR | 8 | 6.06 | 0.19 | FDS+FCR+ECR+EDC | 8 | 5.6 | 0.34 |
| EDC+FCR | 9 | 12.38 | 0.19 | FCR | 9 | 13.3 | 0.23 |
| ECR+FDS+FCR+ | 10 | 12.66 | 0.14 | FCR+ECR | 10 | 15.2 | 0.22 |
| EDC+FDS | 11 | 12.67 | 0.33 | FCR+EDC | 11 | 15.4 | 0.22 |
| EDC | 12 | 12.80 | 0.29 | FCR+ECR+EDC | 12 | 17.3 | 0.21 |
| FCR+FDS | 13 | 16.58 | 0.31 | ECR | 13 | 24.6 | 0.10 |
| FDS | 14 | 19.08 | 0.20 | ECR+EDC | 14 | 26.6 | 0.09 |
| FCR | 15 | 20.48 | 0.32 | EDC | 15 | 30.4 | 0.02 |

## 11. Electronic Supplementary Material

Electronic Supplementary Material 1 (ESM1) – Description of acquisition and processing of hand kinematics data (pdf document)

Electronic Supplementary Material 2 (ESM2) – Description of the musculoskeletal model (pdf document)

Electronic Supplementary Material 3 (ESM3) – Muscle activation results for all measured muscles (pdf document)